\documentclass{jaa}
\usepackage{natbib}
\usepackage[colorlinks=true,citecolor=blue]{hyperref}
\usepackage{amsmath}
\usepackage{tikz}
\usepackage{graphicx}
\usepackage[skip=2pt,font=scriptsize]{caption}
\usepackage{caption}
\usepackage{subcaption}

\begin{document}\sloppy
\title{Intra-night optical variability study of a non-jetted narrow-line Seyfert 1 galaxy: 	SDSS J163401.94$+$480940.1}

\author{Vineet Ojha\textsuperscript{1, 2*}}
\affilOne{\textsuperscript{1} Physical Research Laboratory (PRL), Astronomy and Astrophysics Division, Ahmedabad, India - 380 009\\}
\affilTwo{\textsuperscript{2}Aryabhatta Research Institute of observational sciencES (ARIES), Nainital, India - 263001}

\twocolumn[{

\maketitle

\corres{vineetojhabhu@gmail.com, vineetojha@prl.res.in}
\msinfo{----------}      {---------}

\begin{abstract}
SDSS J163401.94$+$480940.2 is a non-jetted radio-loud narrow-line Seyfert 1 (NLSy1) galaxy. Optical monitoring of this object was carried out in two intra-night sessions each $\geq$ 3 hrs with 3.6m DOT. Intra-night optical variability (INOV) characterization is presented for the first time for this source. We have detected an unexpected remarkable flare in one of two monitoring sessions of SDSS J163401.94$+$480940.2, whose rapid brightening phase implied a minute like doubling time of $\sim$ 22 minutes,  thereby approaching to the extremely fast minute like variability, observed from FSRQ PKS 1222$+$21 at 400 GeV.  The detection of a minute-like variability suggests the existence of relativistic jets with a small viewing angle. We briefly discuss the possible mechanisms for the non-detection of relativistic jets in its Very Long Baseline Array (VLBA) observations.
\end{abstract}

}]

\doinum{12.3456/s78910-011-012-3}
\artcitid{\#\#\#\#}
\pgrange{1--}
\setcounter{page}{1}
\lp{1}

\section{Introduction}
\label{section_1.0}
Narrow-line Seyfert 1 (NLSy1) galaxies are a distinctive subclass of Seyfert galaxies with intense multi-wavelength properties. While in optical wave-band both permitted and forbidden emission lines are present in their spectra but the width of their broad component of Balmer emission lines are narrower than those of normal broad-line Seyfert 1 galaxies with the full width at half maximum of the broad component of Balmer emission line (FWHM(H${\beta}$)) less than 2000 km s$^{-1}$~\citep{Osterbrock1985ApJ...297..166O, Goodrich1989ApJ...342..908G}. In addition to the criterion of FWHM(H${\beta}$), two extreme optical characteristics such as relatively weak [O$_{III}$] and strong permitted Fe~{\sc ii} emission lines with [O$_{III}]_{\lambda5007}/H\beta$ $<$ 3 are used to define the NLSy1 galaxy~\citep{Shuder-Osterbrock1981ApJ...250...55S}. Moreover, NLSy1 galaxies (NLSy1s) also display other utmost observational characteristics such as steep soft X-ray spectra, rapid X-ray (in optical sometimes) flux variability, strong soft X-ray excess below 2 keV, and blue-shifted line profile~\citep[e.g.,][]{Brandt1997MNRAS.285L..25B,Leighly1999ApJS..125..297L, Vaughan1999MNRAS.309..113V, Komossa-Meerschweinchen2000A&A...354..411K, Miller2000NewAR..44..539M, Zamanov2002ApJ...576L...9Z, Klimek2004ApJ...609...69K, Leighly2004ApJ...611..107L, Boroson2005AJ....130..381B, Liu2010ApJ...715L.113L, Paliya2013MNRAS.428.2450P, Kshama2017MNRAS.466.2679K, Ojha2019MNRAS.483.3036O, Ojha2020MNRAS.493.3642O, Ojha2020ApJ...896...95O}. It is believed that NLSy1s are young active galactic nuclei (AGNs) and represents an earlier stage in their evolution~\citep[e.g.,][]{Mathur2000MNRAS.314L..17M, Sulentic2000ApJ...536L...5S, Mathur2001NewA....6..321M, Fraix-Burnet2017FrASS...4....1F, Komossa2018rnls.confE..15K, Paliya2019JApA...40...39P}. Observational evidence suggests that the average estimated black hole mass of NLSy1s based upon various methods such as reverberation mapping, luminosity-radius relationship, and single-epoch virial method, has been found to be relatively lower $\sim 10^{6} - 10^{7} M_{\odot}$  ~\citep{Grupe2004ApJ...606L..41G, Deo2006AJ....132..321D, Zhou2006ApJS..166..128Z, Peterson2011nlsg.confE..32P, Wang2014ApJ...793..108W, Rakshit2017ApJS..229...39R}, and they accrete with a high fraction of the Eddington rate, in contrast to quasars~\citep{Boroson1992ApJS...80..109B, Peterson2000ApJ...542..161P}. However, a systematic underestimation of their black hole masses is suggested~\citep{Decarli2008MNRAS.386L..15D, Marconi2008ApJ...678..693M, Calderone2013MNRAS.431..210C, Viswanath2019ApJ...881L..24V, Ojha2020ApJ...896...95O}. NLSy1s are mainly hosted by spiral/disc galaxies~\citep{Deo2006AJ....132..321D, Ohta2007ApJS..169....1O, 2020MNRAS.492.1450O}. However, elliptical hosts are also suggested for a few $\gamma$-ray detected NLSy1s~\citep[e.g., see ][]{D'Ammando2017MNRAS.469L..11D, D'Ammando2018MNRAS.478L..66D}.\par
Conventionally, it was thought that NLSy1s are often being radio-quiet, usually defined with the ratio (R) of rest-frame flux densities at 5 GHz and 4400\AA~to be $\leq$ 10 ~\citep[see ][and references therein]{Kellermann1994AJ....108.1163K, Kellermann1989AJ.....98.1195K, Kellermann2016ApJ...831..168K}. However, with the recent statistically large sample of NLSy1s, about $\sim$ 7\% of NLSy1s are found to be radio-loud having R $>$ 10~\citep[hereafter RLNLSy1s,][]{Komossa2006AJ....132..531K, Zhou2006ApJS..166..128Z, Rakshit2017ApJS..229...39R, Singh2018MNRAS.480.1796S}, which suggests that in a few of them, jets are present~\citep{Zhou2003ApJ...584..147Z,Yuan2008ApJ...685..801Y}. In fact, parsec-scale blazar-like radio jets were revealed  in the Very Long Baseline Array (VLBA) observations of several NLSy1s~\citep{Lister2013AJ....146..120L, Gu2015ApJS..221....3G, Lister2016AJ....152...12L}. The launching of relativistic jets in a subclass of AGN having smaller black hole masses and higher accretion rates counters the historical trend of the launching of relativistic jets with larger black hole masses and lower accretion rates~\citep{Urry2000ApJ...532..816U, Boroson2002ApJ...565...78B, 2002ApJ...564...86B, Urry2003ASPC..290....3U, Marscher2009arXiv0909.2576M, Chiaberge2011MNRAS.416..917C}, and also challenges the theoretical scenarios of jet formation~\citep[e.g., ][]{2002ApJ...564...86B}. Furthermore, in the case of stellar-mass black holes with high accretion rates generally accord to quenched states of jets~\citep{Boroson2002ApJ...565...78B, Maccarone2003MNRAS.345L..19M}. Therefore, studying jet related aspects of the NLSy1 is important to understand  physical processes that are capable to launch relativistic jets in this unique class of AGN.\par
On the other hand, in addition to a similar double-humped spectral energy distribution (SED) of some RLNLSy1s with blazars~\citep[e.g.,][]{Abdo2009ApJ...707L.142A, Paliya2013ApJ...768...52P, Paliya2019JApA...40...39P}, a significant fraction of RLNLSy1s especially very radio-loud (R $>$ 100) display blazar-like characteristics such as rapid infrared and X-ray flux variability~\citep{Boller1996A&A...305...53B, Grupe1998A&A...330...25G, Leighly1999ApJS..125..297L, Hayashida2000NewAR..44..419H, Komossa-Meerschweinchen2000A&A...354..411K, Jiang2012ApJ...759L..31J,Itoh2013ApJ...775L..26I, Yao2015MNRAS.454L..16Y, Gabanyi2018rnls.confE..42G}, compact radio cores, high brightness temperature, superluminal motion, flat radio and X-ray spectra~\citep{Yuan2008ApJ...685..801Y, Orienti2012arXiv1205.0402O, Berton2018A&A...614A..87B, Lister2018rnls.confE..22L}. All these characteristics give indirect evidence of the presence of jets in them and detections of $\gamma$-ray emissions by {\it Fermi}-Large Area Telescope ({\it Fermi}-LAT)\footnote{https://heasarc.gsfc.nasa.gov/docs/heasarc/missions/fermi.html} from a handful of RLNLSy1s support the scenario that these jets are relativistic~\citep{Abdo2009ApJ...699..976A, Abdo2009ApJ...707..727A, Abdo2009ApJ...707L.142A, Foschini2010ASPC..427..243F, Foschini2011nlsg.confE..24F, D'Ammando2012MNRAS.426..317D, D'Ammando2015MNRAS.452..520D, Yao2015MNRAS.454L..16Y, Paliya2018ApJ...853L...2P, Yang2018MNRAS.477.5127Y, Yao2019MNRAS.487L..40Y}. \par
Flux variability of AGNs on minutes to hour time scales in the optical waveband is termed as intra-night optical variability~\citep[INOV,][]{Gopal-Krishna1993A&A...271...89G}, and this alternative tool is also used to indirectly verify the presence of jets, as it has been observationally well established that radio-loud jet dominated sources such as blazars exhibit a distinctive stronger INOV, both in INOV amplitude ($\psi$) and duty cycle (DC, fractional time for which an AGN is found to be variable) as compared to their radio-quiet counterparts i.e. QSOs. In fact, this tool has been used for a decade as a diagnostic to search for the Doppler boosted optical jets in X-ray detected NLSy1s, $\gamma$-ray detected NLSy1s, and weak emission line QSOs~\citep[e.g., see][]{Liu2010ApJ...715L.113L, Paliya2013MNRAS.428.2450P, Kumar2015MNRAS.448.1463K, Kumar2016MNRAS.461..666K, Kumar2017MNRAS.471..606K, Ojha2018BSRSL..87..387O, Ojha2019MNRAS.483.3036O,  Ojha2021MNRAS.501.4110O}. Therefore, to continue these variability studies, we present here the intra-night variability study of a RLNLSy1 galaxy ($R=204$) SDSS J163401.94$+$480940.2, which we observed with 3.6m Devasthal Optical Telescope  (DOT) of the Aryabhatta Research Institute of   observational SciencES (ARIES), India. The RLNLSy1 SDSS J163401.94$+$480940.2 discussed in this paper is from the member of eight non-jetted NLSy1s (Ojha et al. 2021 under preparation). Out of these  eight NLSy1s, six were already reported (see sample section of Ojha et al. 2021 under preparation for more detail) in~\citet{Gu2015ApJS..221....3G} where they have  confirmed no-jet in them based upon their VLBA observations.\par
The layout of this paper is as follows. A brief introduction about the source is presented in Sect.~\ref{section_source}. Sect.~\ref{section_2.0} provides details of our photometric monitoring and data reduction procedure. The statistical method is presented in Sect.~\ref{section_3.0}. Our main results followed by discussion are given in Sect.~\ref{section_4.0}. 
\section{SDSS J163401.94 + 480940.2}
\label{section_source}
The rather narrower FWHM of H${\beta}$ about 1609$\pm$79 km s$^{-1}$ of SDSS J163401.94$+$480940.2 resulted in a smaller black hole mass of $\sim$  $2.5\times10^{7}$ M$_{\odot}$ based upon single-epoch optical spectroscopy virial method~\citep[see ][]{Yuan2008ApJ...685..801Y}.  In addition to narrower FWHM(H${\beta}$), the small flux ratio [O$_{III}]_{\lambda5007}/H\beta$ of 0.3 and strong permitted Fe~{\sc ii} emission line make it conventional NLSy1~\citep{Yuan2008ApJ...685..801Y}.  It is a radio-loud NLSy1 at $z$ = 0.49 with  R$_{1.4~GHz}$ = 204~\citep{Gu2015ApJS..221....3G}. Its radio spectrum between the Westerbork
Northern Sky Survey (WENSS) 325 MHz and the NRAO VLA Sky Survey (NVSS) 1.4 GHz data of similar resolution has been found to be flat with $\alpha = -0.47$ and its brightness temperature is estimated to be  $10^{10.1}$K  from its high-resolution VLBA image~\citep{Gu2015ApJS..221....3G}. Only a compact component was detected in its high-resolution VLBA image~\citep[see figure 14 of ][]{Gu2015ApJS..221....3G}, therefore,~\citet{Gu2015ApJS..221....3G} have classified this source as a non-jetted source.

\section{Photometric monitoring and data reduction}
\label{section_2.0}

Intranight monitoring of target source SDSS J163401.94$+$480940.1 was carried out in the Bessel broad-band filter R in two epochs each $\geq$ 3.0 hrs with 3.6m DOT of the ARIES, located at Devasthal, India~\citep{Sagar2012SPIE.8444E..1TS}. DOT has a Ritchey–Chretien design, with an f/9 beam at the Cassegrain focus, and an alt-azimuth mounting. Our observations were performed with a 4k$\times$4k CCD imager mounted on the main port of the telescope~\citep[see ][]{Pandey2018BSRSL..87...42P}. The 4k$\times$4k CCD is cooled with liquid nitrogen (LN$_{2}$) to $-$120$^{\circ}$C having a CCD pixel size of 15 $\mu$m and a plate scale of 0.095 arcsec/pixel, covering a field-of-view (FOV) of $\sim$ 6.52$\times$6.52 arcmin$^{2}$ on the sky. Our observations were taken in 2$\times$2 binning mode with a readout speed of 1 MHz which corresponds to the system rms noise and gain of 8.0 $e^-$ and 2.0 $e^-$ ADU$^{-1}$, respectively.\par
For each night, sky flat-field images were taken during dusk and dawn and at least three bias frames were taken in each night. The dark frames were not taken during our observations due to a relatively low temperature (of about $-120^\circ$C) of the CCD detector used. Preliminary processing of the observed frames was done following the standard routines within the {\sc IRAF}\footnote{Image Reduction and Analysis Facility (http://iraf.noao.edu/)} software package. Since the optical field of monitored NLSy1 was not crowded and target object is a point-like without extended emission, therefore, aperture photometry~\citep{1987PASP...99..191S, 1992ASPC...25..297S} was done for extracting the instrumental magnitudes of the target and the comparison stars recorded in the CCD frames, using DAOPHOT II algorithm\footnote{Dominion Astrophysical Observatory Photometry (http://www.astro.wisc.edu/sirtf/daophot2.pdf)}. Aperture size is a key parameter in the photometry for measuring the instrumental magnitude and the corresponding signal-to-noise ratio (S/N) of the individual photometric data points. In addition to aperture size, caution about seeing disc (FWHM) variation during an intra-night session becomes very important for the nearby AGNs because a significant contribution to the total flux can come from the underlying host galaxy. Thus, the relative contributions of the (point-like) AGN and the host galaxy to the aperture photometry can vary significantly as the point spread function (PSF) changes during the session. As a result, in the standard analysis of the differential light curves (DLCs), it might lead to statistically significant, yet spurious claims of INOV for small apertures comparable to the PSF~\citep[see][]{Cellone2000AJ....119.1534C}. Therefore, data reduction, aperture selection, and caution for seeing  disc (FWHM) variations were done following the procedure adopted in~\citet{Ojha2021MNRAS.501.4110O}. 

\begin{figure}
	\begin{center}
	\includegraphics[width=8.0cm, height=7.4cm]{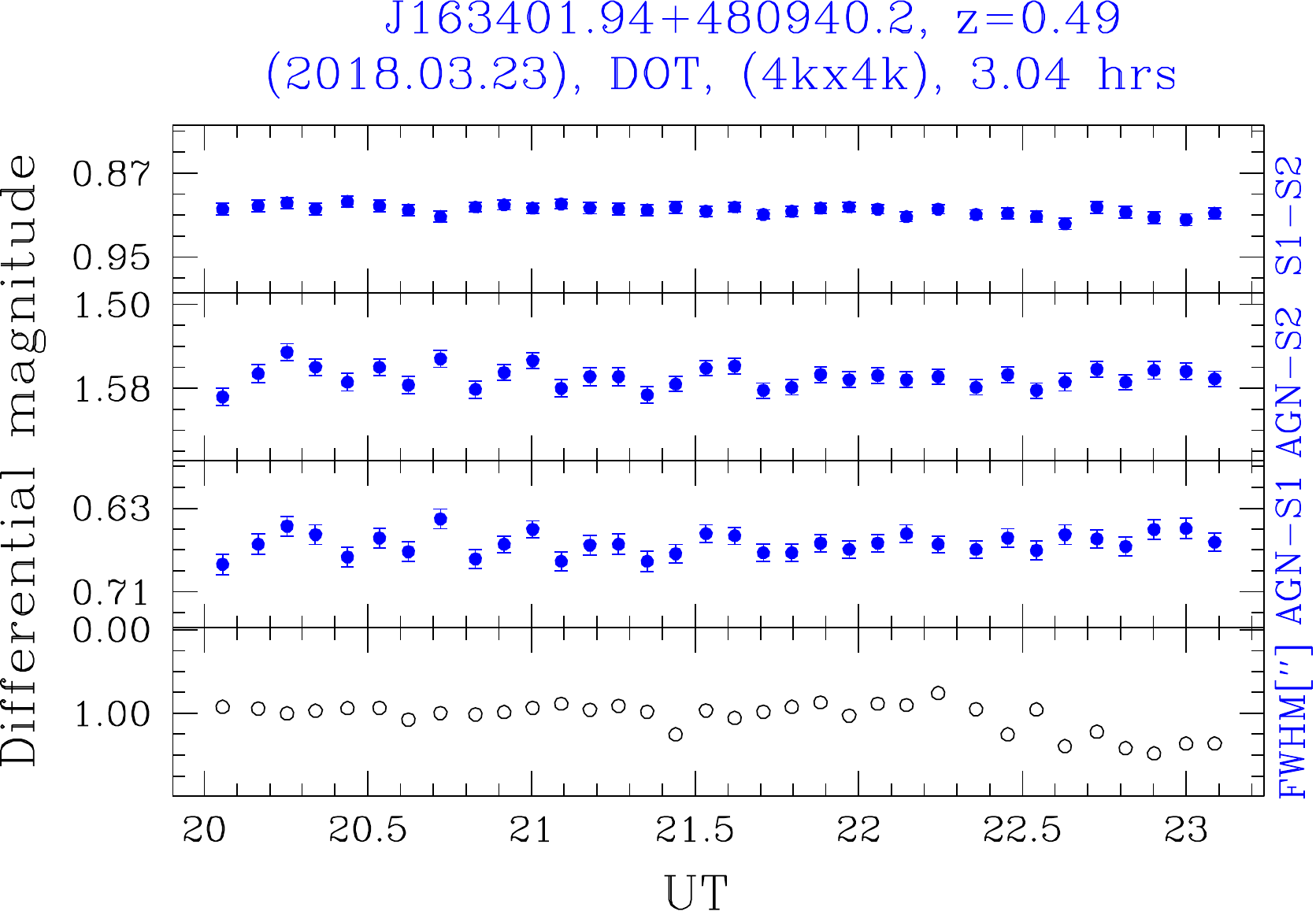}
	\end{center}
	\caption{\small Intranight differential light curves (DLCs) of SDSS J163401.94$+$480940.2. In each panel, the upper DLC is derived using the chosen two (non-varying) comparison stars, while the lower two DLCs are the `RLNLSy1-star' DLCs, as defined  in the labels on the right side. The bottom panel displays the variations of the seeing disc (FWHM) during the monitoring session. The redshift, date of observation, telescope, CCD used, and monitoring time are given at the top of panel.}
	\label{fig:J1634}
\end{figure}

\begin{figure*}
\begin{subfigure}[t]{.5\textwidth}
     \includegraphics[width=8.0cm, height=8.5cm]{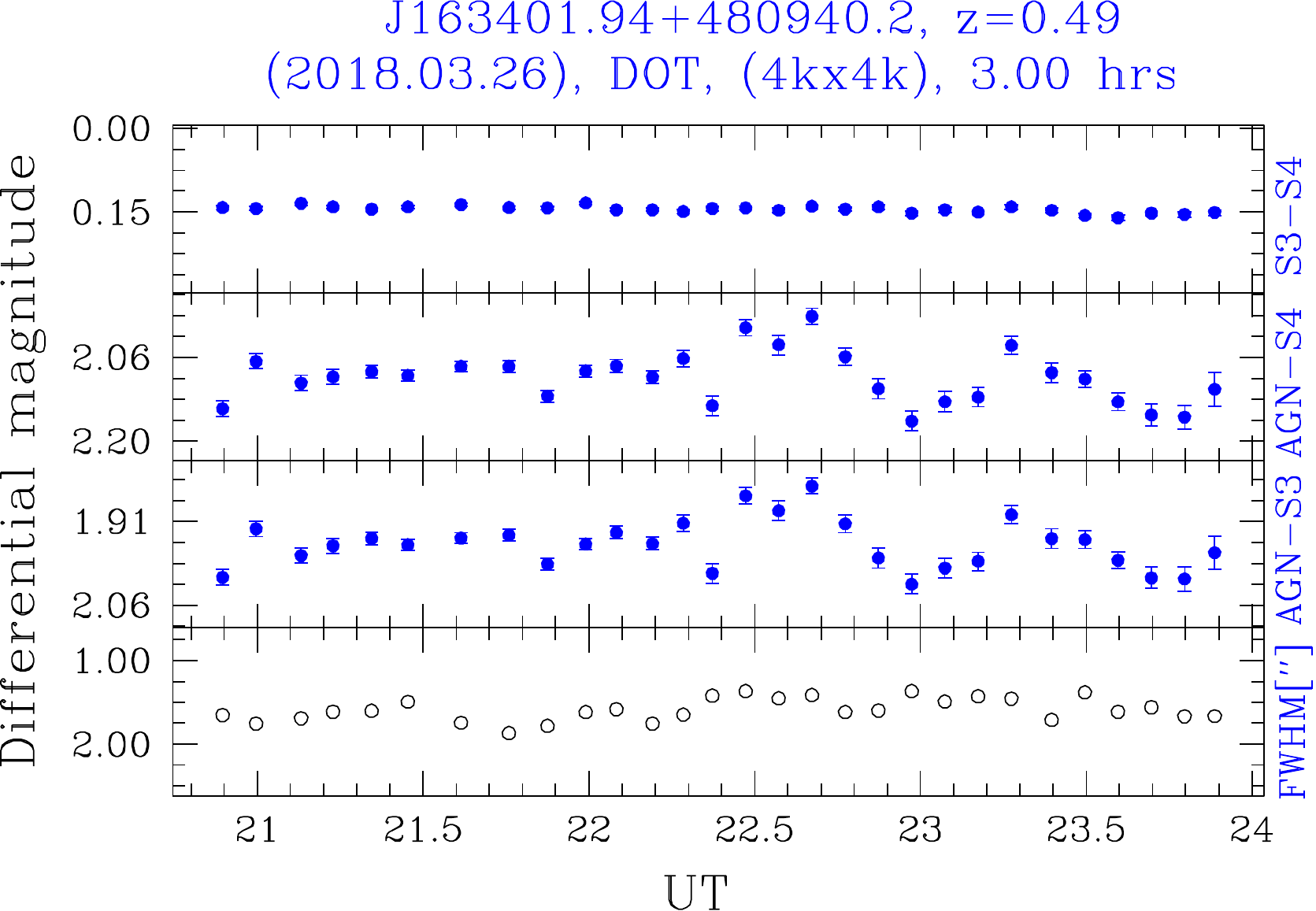}
  \caption{\small A 300s sampled DLCs.}
  \label{fig:sub1}
\end{subfigure}%
\begin{subfigure}[t]{.5\textwidth}
  \includegraphics[width=8.0cm, height=7.0cm]{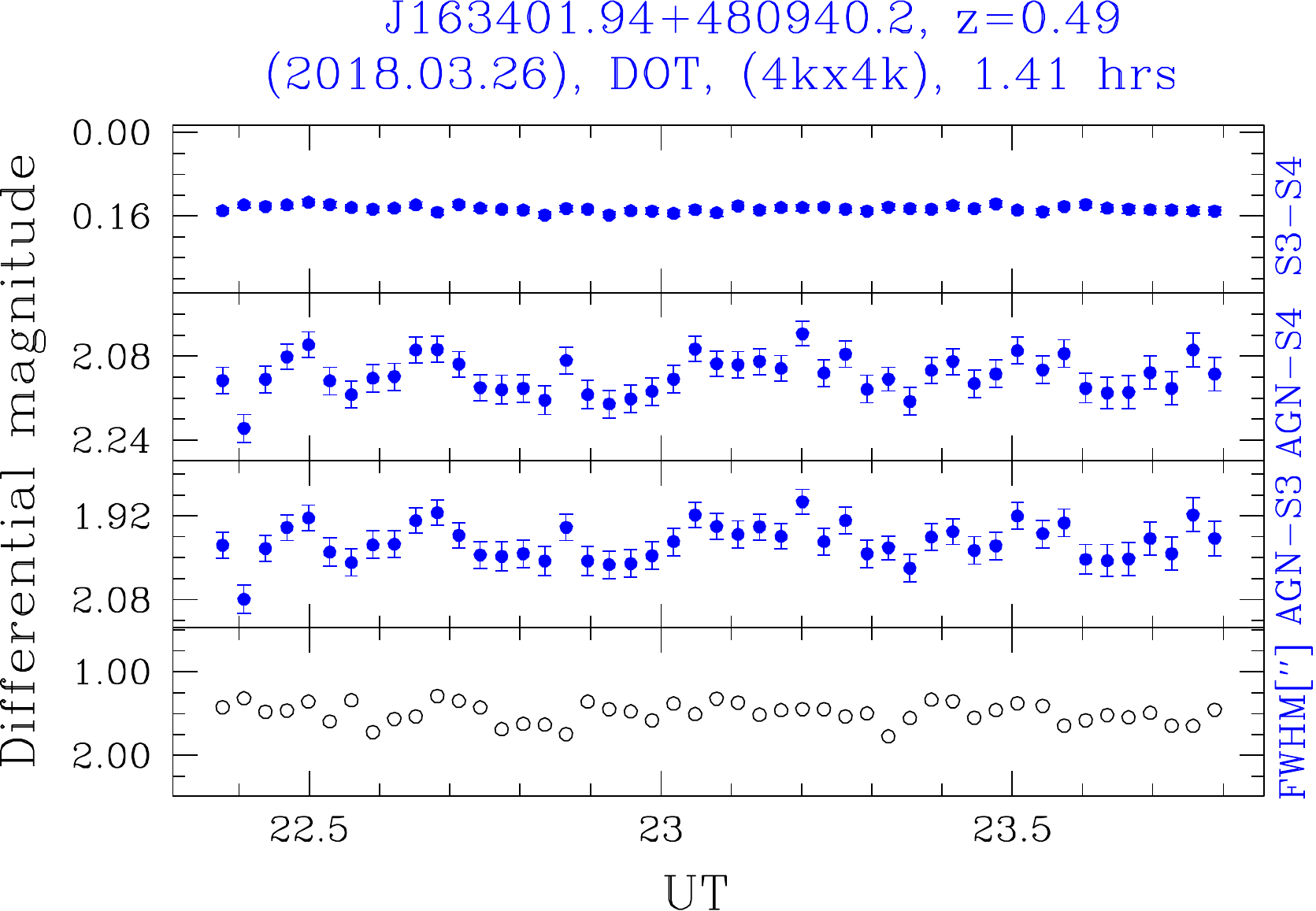}
  \caption{\small A 100s sampled DLCs from flaring point to end point.}
  \label{fig:sub2}
\end{subfigure}
\caption{\small Intranight differential light curves (DLCs) of SDSS J163401.94$+$480940.2 monitored on 2018.03.26, (a): 300 seconds sampled DLCs for whole monitoring session, (b): 100s sampled DLCs of same epoch from flaring point (22.37 UT) to end point (23.88 UT). The panels detail are same as Fig.~\ref{fig:J1634}. }
\label{fig:J1634_flare}
\end{figure*}

\section{Statistical method}
\label{section_3.0}

Since the prime interest of this work is to search INOV in the observed target, therefore, the differential photometric technique is used to produce differential light curves of each observed night following the procedure described in~\citet{Ojha2021MNRAS.501.4110O}. Furthermore,  
for the confirmation of INOV in DLCs of each observed night statistically, we have applied two different versions of the \emph{F$-$test} proposed by~\citet{Diego2010AJ....139.1269D}.  These two tests are known as standard \emph{F$-$test}~\citep[hereafter $F^{\eta}$-test, e.g., see][]{Goyal2012A&A...544A..37G} and the power-enhanced \emph{$F-$test}~\citep[hereafter $F_{enhced}$-test, e.g., see][]{Diego2014AJ....148...93D}. A comprehensive explanation about these two tests is demonstrated in our old papers~\citep[][and references therein]{ Ojha2020MNRAS.493.3642O, Ojha2021MNRAS.501.4110O}.\par

In short, following~\citet{Goyal2012A&A...544A..37G}, $F^{\eta}$-test can be written as

 \begin{equation}
 \label{euafeta}
 F_{1}^{\eta} = \frac{Var_{(NLSy1-cs1)}} {\eta^2 \langle \sigma_{NLSy1-cs1}^2 \rangle},  
 \hspace{0.3cm} F_{2}^{\eta} = \frac{Var_{(NLSy1-cs2)}} { \eta^2 \langle \sigma_{NLSy1-cs2}^2 \rangle}  
 \end{equation}

where $Var_{(NLSy1-cs1)}$ and $Var_{(NLSy1-cs2)}$ are the variances with $\langle \sigma_{NLSy1-cs1}^2 \rangle=\sum_ {i=1}^{N}\sigma^2_{i,~err}(NLSy1-cs1)/N$ and $\langle \sigma_{NLSy1-cs2}^2 \rangle$ being the mean square (formal) rms errors of the individual data points in the `target NLSy1 - comparison star1' and `target NLSy1 - comparison star2' DLCs, respectively.  Here, ``$\eta$'' an error scaling factor is taken to be $1.5$~\citep[see][]{Goyal2012A&A...544A..37G, Ojha2021MNRAS.501.4110O}.
Two critical significance levels, $\alpha= 0.01$ and $ \alpha = 0.05$ that correspond to the confidence levels of 99\% and 95\%, respectively are set by us in the present work. The estimated values of $F^{\eta}$ employing Eq.~\ref{euafeta}  were compared with its adopted critical $F$-values ($F_{c}$), and the SDSS J163401.94$+$4809 is considered to be variable only when the $F^{\eta}$-values computed for both its DLCs are found to be greater than its critical value at 99\% confidence level (hereafter, $F_{c}(0.99)$). In columns 6 and 7 of Table~\ref{NLSy1:tab_result}, we have tabulated the estimated $F^{\eta}$-values and the correspondingly inferred variability status of our target source SDSS J163401.94$+$4809 for its two intra-night sessions.\par

\begin{table*}
  \centering
  \begin{minipage}{190mm}
 \begin{center}   
 {\small
   \caption[caption]{Observational details and the inferred INOV status for the target SDSS J163401.94$+$480940.2 (photometric aperture radius used for analysis = 2$\times$FWHM).}
  \label{NLSy1:tab_result}
 \begin{tabular}{ccc ccccc ccccccc}
   \hline
   {RLNLSy1s} & Date(s)$^{a}$ &  T$^{b}$  & N$^{c}$  & Median$^{d}$ & {$F^{\eta}$-test} & {INOV} & {$F_{enh}$-test} & {INOV} &{$\sqrt { \langle \sigma^2_{i,err} \rangle}$} & $\overline\psi^{g}_{s1, s2}$\\
   (SDSS name) & yyyy.mm.dd & (hrs) & & FWHM  & {$F_1^{\eta}$},{$F_2^{\eta}$} & status$^{e}$ & $F_{enh}$ & status$^{f}$  & (AGN-s)$^{g}$ & (\%) \\
   &&&& (arcsec) &           &{99\%}& & {99\%}&&\\
   {(1)}&{(2)} & {(3)} & {(4)} & {(5)} & {(6)} & {(7)} & {(8)} & {(9)} & {(10)} & {(11)} \\
\hline
 
 J163401.94$+$480940.2 & 2018.03.23 & 3.04 & 34& 0.98& 00.62, 00.71 &  NV, NV  & 01.61 &   NV & 0.012 &    --\\
                       & 2018.03.26 & 3.00 & 29& 1.56& 03.49, 03.83 &  V ,  V  & 03.50 &   V  & 0.023 &   17.50\\

 \hline
  \multicolumn{15}{l}{$^{a}$Date(s) of the monitoring session(s).  $^{b}$Duration of the monitoring session in the observed frame. $^{c}$Number of data points in the DLCs}\\
  \multicolumn{15}{l}{of the monitoring session. $^{d}$Median seeing (FWHM in arcsec) for the session. $^{e, f}$INOV status inferred from F$^{\eta}$ and F$_{enh}$ tests, with V =}\\
  \multicolumn{15}{l}{variable, i.e. confidence level $\geq$ 99\%; and  NV = non-variable, i.e. confidence level $<$ 95\%. $^{g}$Mean amplitude of variability in the two}\\
\multicolumn{15}{l}{DLCs of the target NLSy1 (i.e., relative to the two comparison stars).}
    \end{tabular}  
 }              
\end{center}
 \end{minipage} 
    \end{table*}

The second version of  $F$-test employed in this study is $F_{enhced}$-test. As described in~\citet{Ojha2020MNRAS.493.3642O} $F_{enhced}$-test can be written as

\begin{equation}
\label{Fenh_eq}  
F_{{\rm enhance}} = \frac{s_{{\rm NLSy1}}^2}{s_{\rm stc}^2}, \hspace{0.1cm} s_{\rm stc}^2=\frac{1}{(\sum _{p=1}^k T_p) - k}\sum _{p=1}^{k}\sum _{i=1}^{T_p}s_{p,i}^2
\end{equation}

where $s_{{\rm NLSy1}}^2$ is the variance of the DLC of the target NLSy1 and the reference star (the one proximity in magnitude to the target NLSy1 out of the two selected non-varying comparison stars), while $s_{\rm stc}^2$ is the stacked variance of the DLCs of the comparison stars and the reference star~\citep{Diego2014AJ....148...93D}. $T_{p}$ is the number of observed frames taken for the $p^{th}$ star, and $k$ is the total number of non-varying comparison stars.\par

The scaled square deviation $s_{{\rm p,i}}^2$ defined as

\begin{equation}
\hspace{2.7cm} s_{p,i}^2=\omega _p(m_{p,i}-\bar{m}_{p})^2
\end{equation}

where $m_{p,i}$'s are the differential instrumental magnitudes, and $\bar{m_{p}}$ is the mean differential magnitude of the reference star and the p$^{th}$ comparison star. The scaling factor $\omega_ {p}$~\citep[see also][]{Joshi2011MNRAS.412.2717J} is taken as

\begin{equation}
 \hspace{2.7cm} \omega _p=\frac{\langle\sigma^2_{i,err}(NLSy1-ref)\rangle}{\langle\sigma^2_{i,err}(s_{p}-ref)\rangle}.
  \end{equation}

The $F_{{\rm enhced}}$ value estimated using Eq.~\ref{Fenh_eq} is compared with its $F_{c}(0.99)$ and $F_{c}(0.95)$, and SDSS J163401.94$+$4809 is considered to be variable only when the $F_{{\rm enhced}}$ computed for both its DLCs are found to be greater than its $F_{c}(0.99)$. In columns 8 and 9 of Table~\ref{NLSy1:tab_result}, we have tabulated the estimated values of $F_{{\rm enhced}}$ and the correspondingly inferred variability status of our target source SDSS J163401.94$+$4809 for its two intra-night sessions.\par

Furthermore,  the peak-to-peak amplitude of  INOV ($\psi$) for quantifying the actual variation featured by SDSS J163401.94$+$4809 in its variable session is computed by following the definition given by~\citet{Heidt1996A&A...305...42H}

\begin{equation} 
\hspace{2.5cm} \psi= \sqrt{({P_{max}}-{P_{min}})^2-2\sigma^2}
\end{equation} 

with $P_{min,~max}$ = minimum (maximum) values in the DLC of target NLSy1 relative to comparison stars and $\sigma^2 = \eta^2\langle\sigma^2_{NLSy1-CS}\rangle$, where $\langle\sigma^2_{NLSy1-CS}\rangle$ is the mean square (formal) rms errors of individual data points and $\eta$ = 1.5~\citep[see ][]{Goyal2012A&A...544A..37G}.

\section{Results and discussion}
\label{section_4.0}
The $\gamma$-ray emissions along with the presence of stronger INOV both in amplitude ($\psi$) and duty cycle (DC) suggest the presence of Doppler boosted relativistic jets in the AGNs because of well known beaming effect~\citep[e.g., see Sect.~\ref{section_1.0}, and also][]{Urry1995PASP..107..803U}. Recently~\citet{Ojha2021MNRAS.501.4110O} have reported the INOV characterization of an unbiased sample of 15 $\gamma$-RLNLSy1s, but INOV characteristics of jetted and non-jetted RLNLSy1s are still lacking and poorly known for non-jetted RLNLSy1s. Therefore, in the present work, we report the INOV characterization of a non-jetted RLNLSy1s SDSS J163401.94$+$480940.2 for first the time which was monitored in two observing sessions each $\geq$ 3 hrs with 3.6m DOT.\par
For the unambiguous variability detection of SDSS J163401.94$+$480940.2 in its both observing sessions, we have first selected two non-variable comparison stars in each observing session and then generated DLCs of SDSS J163401.94$+$480940.2 with respect to them. On 2018.03.23, both the selected comparison stars (S1 and S2) were steady (non-variable) during 3.04 hrs of the monitoring, and seeing disc variation was also steady except for a slight variation of about 0.5 arcsec at the end (see Fig.~\ref{fig:J1634}). In this monitoring session, the sampling time was about 300s, and we did not find the source to be variable based upon both tests (see Table~\ref{NLSy1:tab_result}).  Three days later, on 2018.03.26, the source was again monitored for a total duration of 3.0 hrs with a sampling time of 100s. In this monitoring session, the two chosen non-variable comparison stars S1 and S2 of the earlier session become non-steady, therefore, we have chosen two other non-variable comparison stars S3 and S4 for generating DLCs of SDSS J163401.94$+$480940.2 with respect to them. Furthermore, for achieving comparable S/N of DLCs to earlier monitoring session (i.e., 2018.03.23), we have first stacked three 100s frames and thereafter generated the DLCs for the monitoring session of 2018.03.26. On this night, both the comparison stars (S3 and S4) were steady during 3 hrs of the monitoring session, and seeing disc was also steady throughout the session (see Fig.~\ref{fig:J1634_flare} (a)). Unambiguous evidence of blazar like INOV has been detected on this night based on both tests (see Table~\ref{NLSy1:tab_result}).\par
Although, the blazar types optical variability is rarely expected in non-jetted AGNs or misaligned AGNs~\citep{Paliya2013MNRAS.428.2450P, Bhattacharya2019MNRAS.483.3382B}. But, in the present study, we have been found the unexpected remarkable sharp feature in the differential light curves of the non-jetted-RLNLSy1 SDSS J163401.94$+$480940.2 on dated 2018.03.26 with 3.6m DOT. Therefore, we focus here on this sharp feature only. During the 3.0 hrs of continuous monitoring with high sensitivity and about 5 minutes of sampling time, it can be seen in the DLCs of SDSS J163401.94$+$480940.2 that at around \emph{22.37} UT there was a sharp rise (between two consecutive points) of $\sim$ \emph{14}\% within  \emph{6.01} minutes and then, after remaining quiescent for \emph{12} minutes, faded back to almost its initial level (see Fig.~\ref{fig:J1634_flare} (a)). Caution about seeing disc (FWHM) variation during a monitoring session becomes important when AGNs are at small redshift~\citep{Ojha2021MNRAS.501.4110O} because a significant contribution to the total flux can come from the underlying host galaxy and hence the relative contributions of the (point-like) AGN and the host galaxy to the aperture photometry can vary significantly as the PSF changes during the session. This might lead to statistically significant, yet spurious claims of INOV in the standard analysis of DLCs~\citep{Cellone2000AJ....119.1534C}. However, in the present situation, the high redshift $z=0.49$ of the source and a very small around \emph{0.25} arcsec seeing disc variation during the monitoring session (see bottom  panel of Fig.~\ref{fig:J1634_flare} (a)), suggest that this sharp variation (flare) is unlikely to be affected by the host galaxy contribution of this source and seems to be genuine. This is in accord with a recent deep near-infrared imaging study of RLNLSy1s by~\citet{2020MNRAS.492.1450O} using the ESO Very Large Telescope (VLT) from which It can be inferred that any variable contamination arising from the host galaxy can be safely discounted in the case of AGNs at z$\gtrsim$0.5. Nonetheless, the sharp variation in the DLCs of an AGN like we caught in  SDSS J163401.94$+$480940.2 is sometimes suspectable if it is accommodated by just two points. Therefore, we have regenerated DLCs of SDSS J163401.94$+$480940.2 from the flaring point (i.e., at \emph{22.37} UT) to endpoint (i.e., at \emph{23.88} UT) using its 100s sampling time which we had fortunately taken with 3.6m DOT on 2018.03.26 (see above). This extra caution has been taken by us to ensure that whether a flaring event occurred in the SDSS J163401.94$+$480940.2 at \emph{22.37} UT is accommodated by more than two points or not. The regenerated 100s sampled DLCs of SDSS J163401.94$+$480940.2 (see Fig.~\ref{fig:J1634_flare} (b)) has ensured that the flaring event is accommodated by \emph{four} points. This has again ensured that the flaring event discovered in the SDSS J163401.94$+$480940.2 with 3.6m DOT is genuine.\par
Such sharp variations are quite uncommon even in case of $\gamma$-ray detected NLSy1s~\citep[e.g., see][]{Eggen2013ApJ...773...85E, Maune2014ApJ...794...93M, Ojha2019MNRAS.483.3036O} and extremely rare for blazars~\citep[e.g., see][and reference theirin]{Gopal-Krishna2018BSRSL..87..281G}. The high redshift nature of this source ($z=0.49$) implying high intrinsic luminosity capable to swamp the host galaxy allowed us to assert that there is almost no contribution from its host galaxy to our chosen aperture for photometry and most of its optical emissions are due to synchrotron and accretion disc~\citep[e.g., see][]{Ojha2019MNRAS.483.3036O}. Now, if we estimate the flux doubling time on our conservative assumption, that equal contributions i.e. 50\% each is coming from AGN's accretion disc and synchrotron (jet) then in order to account for the observed brightening of $\sim$ \emph{14}\% (in \emph{6.01} minutes), occurred at around \emph{22.37} UT (see Fig.~\ref{fig:J1634_flare} (a)), the optical synchrotron component of AGN is required to have brightened up by a factor of  \emph{1.27}. This corresponds to a flux doubling time of $\sim$ \emph{0.37} hrs ($\sim$ \emph{22} minutes).\par   
The spectacular variation observed in the DLCs of SDSS J163401.94$+$480940.2 is remarkable, as such a variation is genuinely unexpected for the non-jetted-RLNLSy1s. However, exceptional variation, and deduced minutes like flux doubling time support the jet based origin and thereby approaching to the extremely fast minute like variability (with a flux doubling time of $\sim$ 10 minutes), observed from flat spectrum radio quasar (FSRQ) PKS 1222$+$21 at 400 GeV~\citep[e.g., see][]{2011ApJ...730L...8A}. On the other hand, the detection of a compact core component only in the VLBA observation at 5 GHz for this source, classifies it into the non-jetted-RLNLSy1 category~\citep[see][]{Gu2015ApJS..221....3G}, which is contrary to the present findings from this source. This might be either due to the quiescent state of this source during its VLBA observation, performed in 2013~\citep[see][]{Gu2015ApJS..221....3G} or due to limited sensitivity of VLBA. Another reason for non-detection of jet component, in the VLBA observation of this source by~\citet{Gu2015ApJS..221....3G} could be its lower black hole mass of $\sim 2.5\times10^{7}$ M$_{\odot}$ (see Sect.~\ref{section_source}) and lower radio luminosity of $1.02\times10^{41}$ erg s$^{-1}$ at 1.4 GHz~\citep[see table 1 of][]{Gu2015ApJS..221....3G}, indirectly implies to harbor less powerful jet~\citep{Heinz2003MNRAS.343L..59H, Foschini2014IJMPS..2860188F}, hence not capable to escape from the confines of its host galaxy~\citep{Berton2020A&A...636A..64B}. Finally, one last possibility of non-detection of radio jet in the VLBA observation of SDSS J163401.94$+$480940.2, could be either through synchrotron self-absorption as it occurs in gigahertz peaked sources or, more probably, via free-free absorption that can be understood as follows. Since NLSy1s are typically characterized by a high Eddington ratios~\citep{Boroson1992ApJS...80..109B, Peterson2000ApJ...542..161P, Ojha2020ApJ...896...95O}, and they are generally associated with a dense circumnuclear enviroment~\citep{Heckman2014ARA&A..52..589H} with a high star formation activity with respect to regular Seyfert galaxies~\citep{Chen2009ApJ...695L.130C, Sani2010MNRAS.403.1246S}. Therefore, the high star formation along with the nuclear activity of NLSy1s can ionize the circumnuclear gas around it, and thus, the large quantities of ionized gas produced via this process could be responsible for screening the jet emission at low frequency, and hence resulting in non-detection of the jet component at low frequencies observations. Even formation of a cocoon of ionized gas~\citep{Wagner2012ApJ...757..136W, Morganti2017FrASS...4...42M} can also be possible when the jet passes through the interstellar medium, which could also be responsible for the free-free absorption~\citep{Bicknell1997ApJ...485..112B}. 

\section*{Acknowledgements}
I thank an anonymous referee for his/her very important comments that have improved this paper considerably. I am very grateful to Prof. Hum Chand for his helpful discussion and useful suggestions. I also thank the concerned ARIES staff for assistance during the observations obtained at the 3.6m Devasthal Optical Telescope (DOT), which is a National Facility run and managed by ARIES, an autonomous Institute under the Department of Science and Technology, Government of India.

\bibliography{main}

\begin{thebibliography}{}
\expandafter\ifx\csname natexlab\endcsname\relax\def\natexlab#1{#1}\fi

\bibitem[{{Abdo} {$et~al$.}(2009{\natexlab{a}}){Abdo}, {Ackermann}, {Ajello},
  {Axelsson}, {Baldini}, {Ballet}, {Barbiellini}, {Bastieri}, {Battelino},
  {Baughman}, {Bechtol}, {Bellazzini}, {Bloom}, {Bonamente}, {Borgland},
  {Bregeon}, {Brez}, {Brigida}, {Bruel}, {Caliandro}, {Cameron}, {Caraveo},
  {Casandjian}, {Cavazzuti}, {Cecchi}, {Chekhtman}, {Cheung}, {Chiang},
  {Ciprini}, {Claus}, {Cohen-Tanugi}, {Collmar}, {Conrad}, {Costamante},
  {Dermer}, {de Angelis}, {de Palma}, {Digel}, {Silva}, {Drell}, {Dubois},
  {Dumora}, {Farnier}, {Favuzzi}, {Focke}, {Foschini}, {Frailis}, {Fuhrmann},
  {Fukazawa}, {Funk}, {Fusco}, {Gargano}, {Gehrels}, {Germani}, {Giebels},
  {Giglietto}, {Giordano}, {Giroletti}, {Glanzman}, {Grenier}, {Grondin},
  {Grove}, {Guillemot}, {Guiriec}, {Hanabata}, {Harding}, {Hartman},
  {Hayashida}, {Hays}, {Hughes}, {J{\'o}hannesson}, {Johnson}, {Johnson},
  {Johnson}, {Kamae}, {Katagiri}, {Kataoka}, {Kerr}, {Kn{\"o}dlseder}, {Kuehn},
  {Kuss}, {Lande}, {Latronico}, {Lemoine-Goumard}, {Longo}, {Loparco}, {Lott},
  {Lovellette}, {Lubrano}, {Madejski}, {Makeev}, {Max-Moerbeck}, {Mazziotta},
  {McConville}, {McEnery}, {Meurer}, {Michelson}, {Mitthumsiri}, {Mizuno},
  {Monte}, {Monzani}, {Morselli}, {Moskalenko}, {Murgia}, {Nolan}, {Norris},
  {Nuss}, {Ohsugi}, {Omodei}, {Orlando}, {Ormes}, {Paneque}, {Panetta},
  {Parent}, {Pavlidou}, {Pearson}, {Pepe}, {Pesce-Rollins}, {Piron}, {Porter},
  {Rain{\`o}}, {Rando}, {Razzano}, {Readhead}, {Reimer}, {Reimer}, {Reposeur},
  {Richards}, {Ritz}, {Rodriguez}, {Romani}, {Ryde}, {Sadrozinski}, {Sambruna},
  {Sanchez}, {Sander}, {Parkinson}, {Scargle}, {Schalk}, {Sgr{\`o}}, {Smith},
  {Spandre}, {Spinelli}, {Starck}, {Stevenson}, {Strickman}, {Suson},
  {Tagliaferri}, {Takahashi}, {Tanaka}, {Thayer}, {Thompson}, {Tibaldo},
  {Tibolla}, {Torres}, {Tosti}, {Tramacere}, {Uchiyama}, {Usher}, {Vilchez},
  {Vitale}, {Waite}, {Winer}, {Wood}, {Ylinen}, {Zensus}, {Ziegler}, {Fermi/LAT
  Collaboration}, {Ghisellini}, {Maraschi}, {Tavecchio}, \&
  {Angelakis}}]{Abdo2009ApJ...699..976A}
{Abdo}, A.~A., {Ackermann}, M., {Ajello}, M., {$et~al$.} 2009{\natexlab{a}},
  \apj, 699, 976

\bibitem[{{Abdo} {$et~al$.}(2009{\natexlab{b}}){Abdo}, {Ackermann}, {Ajello},
  {Axelsson}, {Baldini}, {Ballet}, {Barbiellini}, {Bastieri}, {Baughman},
  {Bechtol}, \& et~al.}]{Abdo2009ApJ...707..727A}
---. 2009{\natexlab{b}}, \apj, 707, 727

\bibitem[{{Abdo} {$et~al$.}(2009{\natexlab{c}}){Abdo}, {Ackermann}, {Ajello},
  {Baldini}, {Ballet}, {Barbiellini}, {Bastieri}, {Bechtol}, {Bellazzini},
  {Berenji}, {Bloom}, {Bonamente}, {Borgland}, {Bregeon}, {Brez}, {Brigida},
  {Bruel}, {Burnett}, {Caliandro}, {Cameron}, {Caraveo}, {Casandjian},
  {Cecchi}, {{\c C}elik}, {Chekhtman}, {Cheung}, {Chiang}, {Ciprini}, {Claus},
  {Cohen-Tanugi}, {Conrad}, {Cutini}, {Dermer}, {de Palma}, {Silva}, {Drell},
  {Dubois}, {Dumora}, {Farnier}, {Favuzzi}, {Fegan}, {Focke}, {Foschini},
  {Frailis}, {Fukazawa}, {Fusco}, {Gargano}, {Gehrels}, {Germani}, {Giebels},
  {Giglietto}, {Giordano}, {Giroletti}, {Glanzman}, {Godfrey}, {Grenier},
  {Grove}, {Guillemot}, {Guiriec}, {Hayashida}, {Hays}, {Horan}, {Hughes},
  {J{\'o}hannesson}, {Johnson}, {Johnson}, {Kadler}, {Kamae}, {Katagiri},
  {Kataoka}, {Kerr}, {Kn{\"o}dlseder}, {Kuss}, {Lande}, {Latronico}, {Longo},
  {Loparco}, {Lott}, {Lovellette}, {Lubrano}, {Makeev}, {Mazziotta},
  {McConville}, {McEnery}, {Meurer}, {Michelson}, {Mitthumsiri}, {Mizuno},
  {Monte}, {Monzani}, {Morselli}, {Moskalenko}, {Murgia}, {Nolan}, {Norris},
  {Nuss}, {Ohsugi}, {Omodei}, {Orlando}, {Ormes}, {Pelassa}, {Pepe}, {Persic},
  {Pesce-Rollins}, {Piron}, {Porter}, {Rain{\`o}}, {Rando}, {Razzano},
  {Rochester}, {Rodriguez}, {Ryde}, {Sadrozinski}, {Sambruna}, {Sander}, {Saz
  Parkinson}, {Scargle}, {Sgr{\`o}}, {Smith}, {Spandre}, {Spinelli},
  {Strickman}, {Suson}, {Tagliaferri}, {Takahashi}, {Takahashi}, {Tanaka},
  {Thayer}, {Thayer}, {Thompson}, {Tibaldo}, {Tibolla}, {Torres}, {Tosti},
  {Tramacere}, {Uchiyama}, {Usher}, {Vasileiou}, {Vilchez}, {Vitale}, {Waite},
  {Wang}, {Winer}, {Wood}, {Ylinen}, {Ziegler}, {Fermi/LAT Collaboration},
  {Ghisellini}, {Maraschi}, \& {Tavecchio}}]{Abdo2009ApJ...707L.142A}
---. 2009{\natexlab{c}}, \apjl, 707, L142

\bibitem[{{Aleksi{\'c}} {$et~al$.}(2011){Aleksi{\'c}}, {Antonelli}, {Antoranz},
  {Backes}, {Barrio}, {Bastieri}, {Becerra Gonz{\'a}lez}, {Bednarek},
  {Berdyugin}, {Berger}, {Bernardini}, {Biland}, {Blanch}, {Bock}, {Boller},
  {Bonnoli}, {Borla Tridon}, {Braun}, {Bretz}, {Ca{\~n}ellas}, {Carmona},
  {Carosi}, {Colin}, {Colombo}, {Contreras}, {Cortina}, {Cossio}, {Covino},
  {Dazzi}, {De Angelis}, {De Cea del Pozo}, {De Lotto}, {Delgado Mendez},
  {Diago Ortega}, {Doert}, {Dom{\'\i}nguez}, {Dominis Prester}, {Dorner},
  {Doro}, {Elsaesser}, {Ferenc}, {Fonseca}, {Font}, {Fruck}, {Garc{\'\i}a
  L{\'o}pez}, {Garczarczyk}, {Garrido}, {Giavitto}, {Godinovi{\'c}}, {Hadasch},
  {H{\"a}fner}, {Herrero}, {Hildebrand }, {H{\"o}hne-M{\"o}nch}, {Hose},
  {Hrupec}, {Huber}, {Jogler}, {Klepser}, {Kr{\"a}henb{\"u}hl}, {Krause}, {La
  Barbera}, {Lelas}, {Leonardo}, {Lindfors}, {Lombardi}, {L{\'o}pez}, {Lorenz},
  {Makariev}, {Maneva}, {Mankuzhiyil}, {Mannheim}, {Maraschi}, {Mariotti},
  {Mart{\'\i}nez}, {Mazin}, {Meucci}, {Miranda}, {Mirzoyan}, {Miyamoto},
  {Mold{\'o}n}, {Moralejo}, {Nieto}, {Nilsson}, {Orito}, {Oya}, {Paneque},
  {Paoletti}, {Pardo}, {Paredes}, {Partini}, {Pasanen}, {Pauss},
  {Perez-Torres}, {Persic}, {Peruzzo}, {Pilia}, {Pochon}, {Prada}, {Prada
  Moroni}, {Prandini}, {Puljak}, {Reichardt}, {Reinthal}, {Rhode}, {Rib{\'o}},
  {Rico}, {R{\"u}gamer}, {Saggion}, {Saito}, {Saito}, {Salvati}, {Satalecka},
  {Scalzotto}, {Scapin}, {Schultz}, {Schweizer}, {Shayduk}, {Shore},
  {Sillanp{\"a}{\"a}}, {Sitarek}, {Sobczynska}, {Spanier}, {Spiro}, {Stamerra},
  {Steinke}, {Storz}, {Strah}, {Suri{\'c}}, {Takalo}, {Tavecchio}, {Temnikov},
  {Terzi{\'c}}, {Tescaro}, {Teshima}, {Thom}, {Tibolla}, {Torres}, {Treves},
  {Vankov}, {Vogler}, {Wagner}, {Weitzel}, {Zabalza}, {Zandanel}, {Zanin},
  {MAGIC Collaboration}, {Tanaka}, {Wood}, \& {Buson}}]{2011ApJ...730L...8A}
{Aleksi{\'c}}, J., {Antonelli}, L.~A., {Antoranz}, P., {$et~al$.} 2011, \apjl,
  730, L8

\bibitem[{{Berton} {$et~al$.}(2018){Berton}, {Congiu}, {J{\"a}rvel{\"a}},
  {Antonucci}, {Kharb}, {Lister}, {Tarchi}, {Caccianiga}, {Chen}, {Foschini},
  {L{\"a}hteenm{\"a}ki}, {Richards}, {Ciroi}, {Cracco}, {Frezzato}, {La Mura},
  \& {Rafanelli}}]{Berton2018A&A...614A..87B}
{Berton}, M., {Congiu}, E., {J{\"a}rvel{\"a}}, E., {$et~al$.} 2018, \aap, 614,
  A87

\bibitem[{{Berton} {$et~al$.}(2020){Berton}, {J{\"a}rvel{\"a}}, {Crepaldi},
  {L{\"a}hteenm{\"a}ki}, {Tornikoski}, {Congiu}, {Kharb}, {Terreran}, \&
  {Vietri}}]{Berton2020A&A...636A..64B}
{Berton}, M., {J{\"a}rvel{\"a}}, E., {Crepaldi}, L., {$et~al$.} 2020, \aap,
  636, A64

\bibitem[{{Bhattacharya} {$et~al$.}(2019){Bhattacharya}, {Gulati}, \&
  {Stalin}}]{Bhattacharya2019MNRAS.483.3382B}
{Bhattacharya}, D., {Gulati}, S., \& {Stalin}, C.~S. 2019, \mnras, 483, 3382

\bibitem[{{Bicknell} {$et~al$.}(1997){Bicknell}, {Dopita}, \&
  {O'Dea}}]{Bicknell1997ApJ...485..112B}
{Bicknell}, G.~V., {Dopita}, M.~A., \& {O'Dea}, C. P.~O. 1997, \apj, 485, 112

\bibitem[{{Boller} {$et~al$.}(1996){Boller}, {Brandt}, \&
  {Fink}}]{Boller1996A&A...305...53B}
{Boller}, T., {Brandt}, W.~N., \& {Fink}, H. 1996, \aap, 305, 53

\bibitem[{{Boroson}(2005)}]{Boroson2005AJ....130..381B}
{Boroson}, T. 2005, \aj, 130, 381

\bibitem[{{Boroson}(2002)}]{Boroson2002ApJ...565...78B}
{Boroson}, T.~A. 2002, \apj, 565, 78

\bibitem[{{Boroson} \& {Green}(1992)}]{Boroson1992ApJS...80..109B}
{Boroson}, T.~A., \& {Green}, R.~F. 1992, \apjs, 80, 109

\bibitem[{{B{\"o}ttcher} \& {Dermer}(2002)}]{2002ApJ...564...86B}
{B{\"o}ttcher}, M., \& {Dermer}, C.~D. 2002, \apj, 564, 86

\bibitem[{{Brandt} {$et~al$.}(1997){Brandt}, {Mathur}, \&
  {Elvis}}]{Brandt1997MNRAS.285L..25B}
{Brandt}, W.~N., {Mathur}, S., \& {Elvis}, M. 1997, \mnras, 285, L25

\bibitem[{{Calderone} {$et~al$.}(2013){Calderone}, {Ghisellini}, {Colpi}, \&
  {Dotti}}]{Calderone2013MNRAS.431..210C}
{Calderone}, G., {Ghisellini}, G., {Colpi}, M., \& {Dotti}, M. 2013, \mnras,
  431, 210

\bibitem[{{Cellone} {$et~al$.}(2000){Cellone}, {Romero}, \&
  {Combi}}]{Cellone2000AJ....119.1534C}
{Cellone}, S.~A., {Romero}, G.~E., \& {Combi}, J.~A. 2000, \aj, 119, 1534

\bibitem[{{Chen} {$et~al$.}(2009){Chen}, {Wang}, {Yan}, {Hu}, \&
  {Zhang}}]{Chen2009ApJ...695L.130C}
{Chen}, Y.-M., {Wang}, J.-M., {Yan}, C.-S., {Hu}, C., \& {Zhang}, S. 2009,
  \apjl, 695, L130

\bibitem[{{Chiaberge} \& {Marconi}(2011)}]{Chiaberge2011MNRAS.416..917C}
{Chiaberge}, M., \& {Marconi}, A. 2011, \mnras, 416, 917

\bibitem[{{D'Ammando} {$et~al$.}(2018){D'Ammando}, {Acosta-Pulido}, {Capetti},
  {Baldi}, {Orienti}, {Raiteri}, \& {Ramos
  Almeida}}]{D'Ammando2018MNRAS.478L..66D}
{D'Ammando}, F., {Acosta-Pulido}, J.~A., {Capetti}, A., {$et~al$.} 2018,
  \mnras, 478, L66

\bibitem[{{D'Ammando} {$et~al$.}(2017){D'Ammando}, {Acosta-Pulido}, {Capetti},
  {Raiteri}, {Baldi}, {Orienti}, \& {Ramos
  Almeida}}]{D'Ammando2017MNRAS.469L..11D}
---. 2017, \mnras, 469, L11

\bibitem[{{D'Ammando} {$et~al$.}(2015){D'Ammando}, {Orienti}, {Larsson}, \&
  {Giroletti}}]{D'Ammando2015MNRAS.452..520D}
{D'Ammando}, F., {Orienti}, M., {Larsson}, J., \& {Giroletti}, M. 2015, \mnras,
  452, 520

\bibitem[{{D'Ammando} {$et~al$.}(2012){D'Ammando}, {Orienti}, {Finke},
  {Raiteri}, {Angelakis}, {Fuhrmann}, {Giroletti}, {Hovatta}, {Max-Moerbeck},
  {Perkins}, {Readhead}, {Richards}, {Stawarz}, \&
  {Donato}}]{D'Ammando2012MNRAS.426..317D}
{D'Ammando}, F., {Orienti}, M., {Finke}, J., {$et~al$.} 2012, \mnras, 426, 317

\bibitem[{{de Diego}(2010)}]{Diego2010AJ....139.1269D}
{de Diego}, J.~A. 2010, \aj, 139, 1269

\bibitem[{{de Diego}(2014)}]{Diego2014AJ....148...93D}
---. 2014, \aj, 148, 93

\bibitem[{{Decarli} {$et~al$.}(2008){Decarli}, {Dotti}, {Fontana}, \&
  {Haardt}}]{Decarli2008MNRAS.386L..15D}
{Decarli}, R., {Dotti}, M., {Fontana}, M., \& {Haardt}, F. 2008, \mnras, 386,
  L15

\bibitem[{{Deo} {$et~al$.}(2006){Deo}, {Crenshaw}, \&
  {Kraemer}}]{Deo2006AJ....132..321D}
{Deo}, R.~P., {Crenshaw}, D.~M., \& {Kraemer}, S.~B. 2006, \aj, 132, 321

\bibitem[{{Eggen} {$et~al$.}(2013){Eggen}, {Miller}, \&
  {Maune}}]{Eggen2013ApJ...773...85E}
{Eggen}, J.~R., {Miller}, H.~R., \& {Maune}, J.~D. 2013, \apj, 773, 85

\bibitem[{{Foschini}(2011)}]{Foschini2011nlsg.confE..24F}
{Foschini}, L. 2011, in Narrow-Line Seyfert 1 Galaxies and their Place in the
  Universe, 24

\bibitem[{{Foschini}(2014)}]{Foschini2014IJMPS..2860188F}
{Foschini}, L. 2014, in International Journal of Modern Physics Conference
  Series, Vol.~28, International Journal of Modern Physics Conference Series,
  1460188

\bibitem[{{Foschini} {$et~al$.}(2010){Foschini}, {Fermi/Lat Collaboration},
  {Ghisellini}, {Maraschi}, {Tavecchio}, \&
  {Angelakis}}]{Foschini2010ASPC..427..243F}
{Foschini}, L., {Fermi/Lat Collaboration}, {Ghisellini}, G., {$et~al$.} 2010,
  in Astronomical Society of the Pacific Conference Series, Vol. 427, Accretion
  and Ejection in AGN: a Global View, ed. L.~{Maraschi}, G.~{Ghisellini},
  R.~{Della Ceca}, \& F.~{Tavecchio}, 243--248

\bibitem[{{Fraix-Burnet} {$et~al$.}(2017){Fraix-Burnet}, {Marziani},
  {D'Onofrio}, \& {Dultzin}}]{Fraix-Burnet2017FrASS...4....1F}
{Fraix-Burnet}, D., {Marziani}, P., {D'Onofrio}, M., \& {Dultzin}, D. 2017,
  Frontiers in Astronomy and Space Sciences, 4, 1

\bibitem[{{Gabanyi} {$et~al$.}(2018){Gabanyi}, {Moor}, \&
  {Frey}}]{Gabanyi2018rnls.confE..42G}
{Gabanyi}, K., {Moor}, A., \& {Frey}, S. 2018, in Revisiting Narrow-Line
  Seyfert 1 Galaxies and their Place in the Universe, 42

\bibitem[{{Goodrich} {$et~al$.}(1989){Goodrich}, {Stringfellow}, {Penrod}, \&
  {Filippenko}}]{Goodrich1989ApJ...342..908G}
{Goodrich}, R.~W., {Stringfellow}, G.~S., {Penrod}, G.~D., \& {Filippenko},
  A.~V. 1989, \apj, 342, 908

\bibitem[{{Gopal-Krishna} \& {Wiita}(2018)}]{Gopal-Krishna2018BSRSL..87..281G}
{Gopal-Krishna}, \& {Wiita}, P.~J. 2018, Bulletin de la Societe Royale des
  Sciences de Liege, 87, 281

\bibitem[{{Gopal-Krishna} {$et~al$.}(1993){Gopal-Krishna}, {Wiita}, \&
  {Altieri}}]{Gopal-Krishna1993A&A...271...89G}
{Gopal-Krishna}, {Wiita}, P.~J., \& {Altieri}, B. 1993, \aap, 271, 89

\bibitem[{{Goyal} {$et~al$.}(2012){Goyal}, {Gopal-Krishna}, {Wiita}, {Anupama},
  {Sahu}, {Sagar}, \& {Joshi}}]{Goyal2012A&A...544A..37G}
{Goyal}, A., {Gopal-Krishna}, {Wiita}, P.~J., {$et~al$.} 2012, \aap, 544, A37

\bibitem[{{Grupe} {$et~al$.}(1998){Grupe}, {Beuermann}, {Thomas}, {Mannheim},
  \& {Fink}}]{Grupe1998A&A...330...25G}
{Grupe}, D., {Beuermann}, K., {Thomas}, H.-C., {Mannheim}, K., \& {Fink}, H.~H.
  1998, \aap, 330, 25

\bibitem[{{Grupe} \& {Mathur}(2004)}]{Grupe2004ApJ...606L..41G}
{Grupe}, D., \& {Mathur}, S. 2004, \apjl, 606, L41

\bibitem[{{Gu} {$et~al$.}(2015){Gu}, {Chen}, {Komossa}, {Yuan}, {Shen},
  {Wajima}, {Zhou}, \& {Zensus}}]{Gu2015ApJS..221....3G}
{Gu}, M., {Chen}, Y., {Komossa}, S., {$et~al$.} 2015, \apjs, 221, 3

\bibitem[{{Hayashida}(2000)}]{Hayashida2000NewAR..44..419H}
{Hayashida}, K. 2000, \nar, 44, 419

\bibitem[{{Heckman} \& {Best}(2014)}]{Heckman2014ARA&A..52..589H}
{Heckman}, T.~M., \& {Best}, P.~N. 2014, \araa, 52, 589

\bibitem[{{Heidt} \& {Wagner}(1996)}]{Heidt1996A&A...305...42H}
{Heidt}, J., \& {Wagner}, S.~J. 1996, \aap, 305, 42

\bibitem[{{Heinz} \& {Sunyaev}(2003)}]{Heinz2003MNRAS.343L..59H}
{Heinz}, S., \& {Sunyaev}, R.~A. 2003, \mnras, 343, L59

\bibitem[{{Itoh} {$et~al$.}(2013){Itoh}, {Tanaka}, {Fukazawa}, {Kawabata},
  {Kawaguchi}, {Moritani}, {Takaki}, {Ueno}, {Uemura}, {Akitaya}, {Yoshida},
  {Ohsugi}, {Hanayama}, {Miyaji}, \& {Kawai}}]{Itoh2013ApJ...775L..26I}
{Itoh}, R., {Tanaka}, Y.~T., {Fukazawa}, Y., {$et~al$.} 2013, \apjl, 775, L26

\bibitem[{{Jiang} {$et~al$.}(2012){Jiang}, {Zhou}, {Ho}, {Yuan}, {Wang},
  {Dong}, {Jiang}, {Ji}, \& {Tian}}]{Jiang2012ApJ...759L..31J}
{Jiang}, N., {Zhou}, H.-Y., {Ho}, L.~C., {$et~al$.} 2012, \apjl, 759, L31

\bibitem[{{Joshi} {$et~al$.}(2011){Joshi}, {Chand}, {Gupta}, \&
  {Wiita}}]{Joshi2011MNRAS.412.2717J}
{Joshi}, R., {Chand}, H., {Gupta}, A.~C., \& {Wiita}, P.~J. 2011, \mnras, 412,
  2717

\bibitem[{{Kellermann} {$et~al$.}(2016){Kellermann}, {Condon}, {Kimball},
  {Perley}, \& {Ivezi{\'c}}}]{Kellermann2016ApJ...831..168K}
{Kellermann}, K.~I., {Condon}, J.~J., {Kimball}, A.~E., {Perley}, R.~A., \&
  {Ivezi{\'c}}, {\v Z}. 2016, \apj, 831, 168

\bibitem[{{Kellermann} {$et~al$.}(1989){Kellermann}, {Sramek}, {Schmidt},
  {Shaffer}, \& {Green}}]{Kellermann1989AJ.....98.1195K}
{Kellermann}, K.~I., {Sramek}, R., {Schmidt}, M., {Shaffer}, D.~B., \& {Green},
  R. 1989, \aj, 98, 1195

\bibitem[{{Kellermann} {$et~al$.}(1994){Kellermann}, {Sramek}, {Schmidt},
  {Green}, \& {Shaffer}}]{Kellermann1994AJ....108.1163K}
{Kellermann}, K.~I., {Sramek}, R.~A., {Schmidt}, M., {Green}, R.~F., \&
  {Shaffer}, D.~B. 1994, \aj, 108, 1163

\bibitem[{{Klimek} {$et~al$.}(2004){Klimek}, {Gaskell}, \&
  {Hedrick}}]{Klimek2004ApJ...609...69K}
{Klimek}, E.~S., {Gaskell}, C.~M., \& {Hedrick}, C.~H. 2004, \apj, 609, 69

\bibitem[{{Komossa}(2018)}]{Komossa2018rnls.confE..15K}
{Komossa}, S. 2018, in Revisiting Narrow-Line Seyfert 1 Galaxies and their
  Place in the Universe, 15

\bibitem[{{Komossa} \&
  {Meerschweinchen}(2000)}]{Komossa-Meerschweinchen2000A&A...354..411K}
{Komossa}, S., \& {Meerschweinchen}, J. 2000, \aap, 354, 411

\bibitem[{{Komossa} {$et~al$.}(2006){Komossa}, {Voges}, {Xu}, {Mathur},
  {Adorf}, {Lemson}, {Duschl}, \& {Grupe}}]{Komossa2006AJ....132..531K}
{Komossa}, S., {Voges}, W., {Xu}, D., {$et~al$.} 2006, \aj, 132, 531

\bibitem[{{Kshama} {$et~al$.}(2017){Kshama}, {Paliya}, \&
  {Stalin}}]{Kshama2017MNRAS.466.2679K}
{Kshama}, S.~K., {Paliya}, V.~S., \& {Stalin}, C.~S. 2017, \mnras, 466, 2679

\bibitem[{{Kumar} {$et~al$.}(2016){Kumar}, {Chand}, \&
  {Gopal-Krishna}}]{Kumar2016MNRAS.461..666K}
{Kumar}, P., {Chand}, H., \& {Gopal-Krishna}. 2016, \mnras, 461, 666

\bibitem[{{Kumar} {$et~al$.}(2017){Kumar}, {Gopal-Krishna}, {Stalin}, {Chand},
  {Srianand}, \& {Petitjean}}]{Kumar2017MNRAS.471..606K}
{Kumar}, P., {Gopal-Krishna}, {Stalin}, C.~S., {$et~al$.} 2017, \mnras, 471,
  606

\bibitem[{{Kumar} \& {Gopal-Krishna}and
  {Chand}(2015)}]{Kumar2015MNRAS.448.1463K}
{Kumar}, P., \& {Gopal-Krishna}and {Chand}, H. 2015, \mnras, 448, 1463

\bibitem[{{Leighly}(1999)}]{Leighly1999ApJS..125..297L}
{Leighly}, K.~M. 1999, \apjs, 125, 297

\bibitem[{{Leighly} \& {Moore}(2004)}]{Leighly2004ApJ...611..107L}
{Leighly}, K.~M., \& {Moore}, J.~R. 2004, \apj, 611, 107

\bibitem[{{Lister}(2018)}]{Lister2018rnls.confE..22L}
{Lister}, M. 2018, in Revisiting Narrow-Line Seyfert 1 Galaxies and their Place
  in the Universe, 22

\bibitem[{{Lister} {$et~al$.}(2013){Lister}, {Aller}, {Aller}, {Homan},
  {Kellermann}, {Kovalev}, {Pushkarev}, {Richards}, {Ros}, \&
  {Savolainen}}]{Lister2013AJ....146..120L}
{Lister}, M.~L., {Aller}, M.~F., {Aller}, H.~D., {$et~al$.} 2013, \aj, 146, 120

\bibitem[{{Lister} {$et~al$.}(2016){Lister}, {Aller}, {Aller}, {Homan},
  {Kellermann}, {Kovalev}, {Pushkarev}, {Richards}, {Ros}, \&
  {Savolainen}}]{Lister2016AJ....152...12L}
---. 2016, \aj, 152, 12

\bibitem[{{Liu} {$et~al$.}(2010){Liu}, {Wang}, {Mao}, \&
  {Wei}}]{Liu2010ApJ...715L.113L}
{Liu}, H., {Wang}, J., {Mao}, Y., \& {Wei}, J. 2010, \apjl, 715, L113

\bibitem[{{Maccarone} {$et~al$.}(2003){Maccarone}, {Gallo}, \&
  {Fender}}]{Maccarone2003MNRAS.345L..19M}
{Maccarone}, T.~J., {Gallo}, E., \& {Fender}, R. 2003, \mnras, 345, L19

\bibitem[{{Marconi} {$et~al$.}(2008){Marconi}, {Axon}, {Maiolino}, {Nagao},
  {Pastorini}, {Pietrini}, {Robinson}, \&
  {Torricelli}}]{Marconi2008ApJ...678..693M}
{Marconi}, A., {Axon}, D.~J., {Maiolino}, R., {$et~al$.} 2008, \apj, 678, 693

\bibitem[{{Marscher}(2009)}]{Marscher2009arXiv0909.2576M}
{Marscher}, A.~P. 2009, arXiv e-prints, arXiv:0909.2576

\bibitem[{{Mathur}(2000)}]{Mathur2000MNRAS.314L..17M}
{Mathur}, S. 2000, \mnras, 314, L17

\bibitem[{{Mathur} {$et~al$.}(2001){Mathur}, {Kuraszkiewicz}, \&
  {Czerny}}]{Mathur2001NewA....6..321M}
{Mathur}, S., {Kuraszkiewicz}, J., \& {Czerny}, B. 2001, \na, 6, 321

\bibitem[{{Maune} {$et~al$.}(2014){Maune}, {Eggen}, {Miller}, {Marshall},
  {Readhead}, {Hovatta}, \& {King}}]{Maune2014ApJ...794...93M}
{Maune}, J.~D., {Eggen}, J.~R., {Miller}, H.~R., {$et~al$.} 2014, \apj, 794, 93

\bibitem[{{Miller} {$et~al$.}(2000){Miller}, {Ferrara}, {McFarland}, {Wilson},
  {Daya}, \& {Fried}}]{Miller2000NewAR..44..539M}
{Miller}, H.~R., {Ferrara}, E.~C., {McFarland}, J.~P., {$et~al$.} 2000, \nar,
  44, 539

\bibitem[{{Morganti}(2017)}]{Morganti2017FrASS...4...42M}
{Morganti}, R. 2017, Frontiers in Astronomy and Space Sciences, 4, 42

\bibitem[{{Ohta} {$et~al$.}(2007){Ohta}, {Aoki}, {Kawaguchi}, \&
  {Kiuchi}}]{Ohta2007ApJS..169....1O}
{Ohta}, K., {Aoki}, K., {Kawaguchi}, T., \& {Kiuchi}, G. 2007, \apjs, 169, 1

\bibitem[{{Ojha} {$et~al$.}(2020{\natexlab{a}}){Ojha}, {Chand}, {Dewangan}, \&
  {Rakshit}}]{Ojha2020ApJ...896...95O}
{Ojha}, V., {Chand}, H., {Dewangan}, G.~C., \& {Rakshit}, S.
  2020{\natexlab{a}}, \apj, 896, 95

\bibitem[{{Ojha} {$et~al$.}(2018){Ojha}, {Chand}, \&
  {Gopal-Krishna}}]{Ojha2018BSRSL..87..387O}
{Ojha}, V., {Chand}, H., \& {Gopal-Krishna}. 2018, Bulletin de la Societe
  Royale des Sciences de Liege, 87, 387

\bibitem[{{Ojha} {$et~al$.}(2021){Ojha}, {Chand}, \& {Gopal
  Krishna}}]{Ojha2021MNRAS.501.4110O}
{Ojha}, V., {Chand}, H., \& {Gopal Krishna}. 2021, \mnras, 501, 4110

\bibitem[{{Ojha} {$et~al$.}(2020{\natexlab{b}}){Ojha}, {Chand},
  {Gopal-Krishna}, {Mishra}, \& {Chand}}]{Ojha2020MNRAS.493.3642O}
{Ojha}, V., {Chand}, H., {Gopal-Krishna}, {Mishra}, S., \& {Chand}, K.
  2020{\natexlab{b}}, \mnras, 493, 3642

\bibitem[{{Ojha} {$et~al$.}(2019){Ojha}, {Gopal-Krishna}, \&
  {Chand}}]{Ojha2019MNRAS.483.3036O}
{Ojha}, V., {Gopal-Krishna}, \& {Chand}, H. 2019, \mnras, 483, 3036

\bibitem[{{Olgu{\'\i}n-Iglesias} {$et~al$.}(2020){Olgu{\'\i}n-Iglesias},
  {Kotilainen}, \& {Chavushyan}}]{2020MNRAS.492.1450O}
{Olgu{\'\i}n-Iglesias}, A., {Kotilainen}, J., \& {Chavushyan}, V. 2020, \mnras,
  492, 1450

\bibitem[{{Orienti} {$et~al$.}(2012){Orienti}, {D'Ammando}, {Giroletti}, \&
  {for the Fermi-LAT Collaboration}}]{Orienti2012arXiv1205.0402O}
{Orienti}, M., {D'Ammando}, F., {Giroletti}, M., \& {for the Fermi-LAT
  Collaboration}. 2012, ArXiv e-prints 1205.0402, arXiv:1205.0402

\bibitem[{{Osterbrock} \& {Pogge}(1985)}]{Osterbrock1985ApJ...297..166O}
{Osterbrock}, D.~E., \& {Pogge}, R.~W. 1985, \apj, 297, 166

\bibitem[{{Paliya}(2019)}]{Paliya2019JApA...40...39P}
{Paliya}, V.~S. 2019, Journal of Astrophysics and Astronomy, 40, 39

\bibitem[{{Paliya} {$et~al$.}(2018){Paliya}, {Ajello}, {Rakshit}, {Mandal},
  {Stalin}, {Kaur}, \& {Hartmann}}]{Paliya2018ApJ...853L...2P}
{Paliya}, V.~S., {Ajello}, M., {Rakshit}, S., {$et~al$.} 2018, \apjl, 853, L2

\bibitem[{{Paliya} {$et~al$.}(2013{\natexlab{a}}){Paliya}, {Stalin}, {Kumar},
  {Kumar}, {Bhatt}, {Pandey}, \& {Yadav}}]{Paliya2013MNRAS.428.2450P}
{Paliya}, V.~S., {Stalin}, C.~S., {Kumar}, B., {$et~al$.} 2013{\natexlab{a}},
  \mnras, 428, 2450

\bibitem[{{Paliya} {$et~al$.}(2013{\natexlab{b}}){Paliya}, {Stalin}, {Shukla},
  \& {Sahayanathan}}]{Paliya2013ApJ...768...52P}
{Paliya}, V.~S., {Stalin}, C.~S., {Shukla}, A., \& {Sahayanathan}, S.
  2013{\natexlab{b}}, \apj, 768, 52

\bibitem[{{Pandey} {$et~al$.}(2018){Pandey}, {Yadav}, {Nanjappa}, {Yadav},
  {Reddy}, {Sahu}, \& {Srinivasan}}]{Pandey2018BSRSL..87...42P}
{Pandey}, S.~B., {Yadav}, R. K.~S., {Nanjappa}, N., {$et~al$.} 2018, Bulletin
  de la Societe Royale des Sciences de Liege, 87, 42

\bibitem[{{Peterson}(2011)}]{Peterson2011nlsg.confE..32P}
{Peterson}, B.~M. 2011, in Narrow-Line Seyfert 1 Galaxies and their Place in
  the Universe, 32

\bibitem[{{Peterson} {$et~al$.}(2000){Peterson}, {McHardy}, {Wilkes},
  {Berlind}, {Bertram}, {Calkins}, {Collier}, {Huchra}, {Mathur}, {Papadakis},
  {Peters}, {Pogge}, {Romano}, {Tokarz}, {Uttley}, {Vestergaard}, \&
  {Wagner}}]{Peterson2000ApJ...542..161P}
{Peterson}, B.~M., {McHardy}, I.~M., {Wilkes}, B.~J., {$et~al$.} 2000, \apj,
  542, 161

\bibitem[{{Rakshit} {$et~al$.}(2017){Rakshit}, {Stalin}, {Chand}, \&
  {Zhang}}]{Rakshit2017ApJS..229...39R}
{Rakshit}, S., {Stalin}, C.~S., {Chand}, H., \& {Zhang}, X.-G. 2017, \apjs,
  229, 39

\bibitem[{{Sagar} {$et~al$.}(2012){Sagar}, {Kumar}, {Omar}, \& {Pand
  ey}}]{Sagar2012SPIE.8444E..1TS}
{Sagar}, R., {Kumar}, B., {Omar}, A., \& {Pand ey}, A.~K. 2012, in Society of
  Photo-Optical Instrumentation Engineers (SPIE) Conference Series, Vol. 8444,
  \procspie, 84441T

\bibitem[{{Sani} {$et~al$.}(2010){Sani}, {Lutz}, {Risaliti}, {Netzer}, {Gallo},
  {Trakhtenbrot}, {Sturm}, \& {Boller}}]{Sani2010MNRAS.403.1246S}
{Sani}, E., {Lutz}, D., {Risaliti}, G., {$et~al$.} 2010, \mnras, 403, 1246

\bibitem[{{Shuder} \&
  {Osterbrock}(1981)}]{Shuder-Osterbrock1981ApJ...250...55S}
{Shuder}, J.~M., \& {Osterbrock}, D.~E. 1981, \apj, 250, 55

\bibitem[{{Singh} \& {Chand}(2018)}]{Singh2018MNRAS.480.1796S}
{Singh}, V., \& {Chand}, H. 2018, \mnras, 480, 1796

\bibitem[{{Stetson}(1987)}]{1987PASP...99..191S}
{Stetson}, P.~B. 1987, \pasp, 99, 191

\bibitem[{{Stetson}(1992)}]{1992ASPC...25..297S}
{Stetson}, P.~B. 1992, in Astronomical Society of the Pacific Conference
  Series, Vol.~25, Astronomical Data Analysis Software and Systems I, ed. D.~M.
  {Worrall}, C.~{Biemesderfer}, \& J.~{Barnes}, 297

\bibitem[{{Sulentic} {$et~al$.}(2000){Sulentic}, {Zwitter}, {Marziani}, \&
  {Dultzin-Hacyan}}]{Sulentic2000ApJ...536L...5S}
{Sulentic}, J.~W., {Zwitter}, T., {Marziani}, P., \& {Dultzin-Hacyan}, D. 2000,
  \apjl, 536, L5

\bibitem[{{Urry} \& {Padovani}(1995)}]{Urry1995PASP..107..803U}
{Urry}, C.~M., \& {Padovani}, P. 1995, \pasp, 107, 803

\bibitem[{{Urry} {$et~al$.}(2000){Urry}, {Scarpa}, {O'Dowd}, {Falomo}, {Pesce},
  \& {Treves}}]{Urry2000ApJ...532..816U}
{Urry}, C.~M., {Scarpa}, R., {O'Dowd}, M., {$et~al$.} 2000, \apj, 532, 816

\bibitem[{{Urry}(2003)}]{Urry2003ASPC..290....3U}
{Urry}, M. 2003, in Astronomical Society of the Pacific Conference Series, Vol.
  290, Active Galactic Nuclei: From Central Engine to Host Galaxy, ed.
  S.~{Collin}, F.~{Combes}, \& I.~{Shlosman}, 3

\bibitem[{{Vaughan} {$et~al$.}(1999){Vaughan}, {Reeves}, {Warwick}, \&
  {Edelson}}]{Vaughan1999MNRAS.309..113V}
{Vaughan}, S., {Reeves}, J., {Warwick}, R., \& {Edelson}, R. 1999, \mnras, 309,
  113

\bibitem[{{Viswanath} {$et~al$.}(2019){Viswanath}, {Stalin}, {Rakshit},
  {Kurian}, {Ujjwal}, {Gudennavar}, \& {Kartha}}]{Viswanath2019ApJ...881L..24V}
{Viswanath}, G., {Stalin}, C.~S., {Rakshit}, S., {$et~al$.} 2019, \apjl, 881,
  L24

\bibitem[{{Wagner} {$et~al$.}(2012){Wagner}, {Bicknell}, \&
  {Umemura}}]{Wagner2012ApJ...757..136W}
{Wagner}, A.~Y., {Bicknell}, G.~V., \& {Umemura}, M. 2012, \apj, 757, 136

\bibitem[{{Wang} {$et~al$.}(2014){Wang}, {Du}, {Hu}, {Netzer}, {Bai}, {Lu},
  {Kaspi}, {Qiu}, {Li}, {Wang}, \& {SEAMBH
  Collaboration}}]{Wang2014ApJ...793..108W}
{Wang}, J.-M., {Du}, P., {Hu}, C., {$et~al$.} 2014, \apj, 793, 108

\bibitem[{{Yang} {$et~al$.}(2018){Yang}, {Yuan}, {Yao}, {Li}, {Zhang}, {Zhou},
  {Komossa}, {Liu}, \& {Jin}}]{Yang2018MNRAS.477.5127Y}
{Yang}, H., {Yuan}, W., {Yao}, S., {$et~al$.} 2018, \mnras, 477, 5127

\bibitem[{{Yao} {$et~al$.}(2019){Yao}, {Komossa}, {Liu}, {Yi}, {Yuan}, {Zhou},
  \& {Wu}}]{Yao2019MNRAS.487L..40Y}
{Yao}, S., {Komossa}, S., {Liu}, W.-J., {$et~al$.} 2019, \mnras, 487, L40

\bibitem[{{Yao} {$et~al$.}(2015){Yao}, {Yuan}, {Zhou}, {Komossa}, {Zhang},
  {Qiao}, \& {Liu}}]{Yao2015MNRAS.454L..16Y}
{Yao}, S., {Yuan}, W., {Zhou}, H., {$et~al$.} 2015, \mnras, 454, L16

\bibitem[{{Yuan} {$et~al$.}(2008){Yuan}, {Zhou}, {Komossa}, {Dong}, {Wang},
  {Lu}, \& {Bai}}]{Yuan2008ApJ...685..801Y}
{Yuan}, W., {Zhou}, H.~Y., {Komossa}, S., {$et~al$.} 2008, \apj, 685, 801

\bibitem[{{Zamanov} {$et~al$.}(2002){Zamanov}, {Marziani}, {Sulentic},
  {Calvani}, {Dultzin-Hacyan}, \& {Bachev}}]{Zamanov2002ApJ...576L...9Z}
{Zamanov}, R., {Marziani}, P., {Sulentic}, J.~W., {$et~al$.} 2002, \apjl, 576,
  L9

\bibitem[{{Zhou} {$et~al$.}(2006){Zhou}, {Wang}, {Yuan}, {Lu}, {Dong}, {Wang},
  \& {Lu}}]{Zhou2006ApJS..166..128Z}
{Zhou}, H., {Wang}, T., {Yuan}, W., {$et~al$.} 2006, \apjs, 166, 128

\bibitem[{{Zhou} {$et~al$.}(2003){Zhou}, {Wang}, {Dong}, {Zhou}, \&
  {Li}}]{Zhou2003ApJ...584..147Z}
{Zhou}, H.-Y., {Wang}, T.-G., {Dong}, X.-B., {Zhou}, Y.-Y., \& {Li}, C. 2003,
  \apj, 584, 147

\end{thebibliography}
\bibliographystyle{apj}

\end{document}